\documentclass[preprint,5p,twocolumn]{elsarticle}
\usepackage{graphicx}
\usepackage{float}
\usepackage{amsmath}
\usepackage{amssymb}
\usepackage{color}
\usepackage{tikz}
\newcommand{\vc}[1]{\mathbf{#1}}
\usepackage{subcaption}
\usepackage[utf8]{inputenc}

\journal{Energy Conversion and Management}
\begin{document}

\begin{frontmatter}

\title{Neural networks in day-ahead electricity price forecasting:\\single vs.\ multiple outputs}

\author[wroclaw]{Grzegorz Marcjasz\corref{cor1}}
\ead{grzegorz.marcjasz@pwr.edu.pl}
\author[delft,ev,vito]{Jesus Lago}
\author[wroclaw]{Rafał Weron}

\address[wroclaw]{Department of Operations Research, Wrocław University of Science and Technology, Wrocław, Poland}
\address[delft]{Delft Center for Systems and Control, Delft University of Technology, Delft, The Netherlands}
\address[ev]{Algorithms, Modeling, and Optimization, Energyville, Genk, Belgium}
\address[vito]{Energy Technology, Flemish Institute for Technological Research (VITO), Mol, Belgium}

\cortext[cor1]{Corresponding author}
\begin{abstract}
Recent advancements in the fields of artificial intelligence and machine learning methods resulted in a significant increase of their popularity in the literature, including electricity price forecasting. Said methods cover a very broad spectrum, from decision trees, through random forests to various artificial neural network models and hybrid approaches. In electricity price forecasting, neural networks are the most popular machine learning method as they provide a non-linear counterpart for well-tested linear regression models. Their application, however, is not straightforward, with multiple implementation factors to consider. One of such factors is the network's structure. This paper provides a comprehensive comparison of two most common structures when using the deep neural networks -- one that focuses on each hour of the day separately, and one that reflects the daily auction structure and models vectors of the prices. The results show a significant accuracy advantage of using the latter, confirmed on data from five distinct power exchanges.
\end{abstract}

\end{frontmatter}

\section{Introduction}
The electricity price characteristics are undergoing a process of continuous changes, with the main driving factors being the ever-growing demand and the very eager introduction of the renewable energy sources into the generation mix \cite{gre:vas:10,bra:bri:hod:16}. This results in much more volatile price curves with price spikes observed more often than in the past, and increasingly often reaching negative values. As an consequence, the need for accurate price forecasts is higher than ever before. Utility companies that strongly relies on the electricity prices (e.g., a power plant) can incorporate the information into everyday planning, and as a result -- transform the better forecasts into optimal schedule of the production. The market participants -- and the EPF literature, see e.g. \cite{nar:zie:19,mar:uni:wer:20} -- exhibit increased interest in the intraday market that allows for greater flexibility than the day-ahead, as the exact quantities produced need not to be known the day before the delivery, but rather hours to even minutes before. However, the vast majority of electricity is still traded in the day-ahead markets, even in Germany, where the intraday market adoption is among the highest in Europe and grows continuously \cite{epex:18}.

An increase of the interest in the artificial intelligence (AI) methods has been observed in many research fields, including electricity price forecasting. Several new methods and approaches have been proposed in the last few years, ranging from the most basic machine learning (ML) methods, through parsimonious multilayer perceptron (MLP) neural networks, to sophisticated hybrid or multi-stage solutions \cite{wer:14,mar:uni:wer:19,lag:rid:sch:18,wan:zha:che:16}. Two common factors can be observed in the papers: \emph{i)} ML methods tend to significantly outperform different solutions, such as statistical methods and \emph{ii)} the application of ML methods  requires the user to provide multiple hyperparameter values and to decide on a range of additional aspects of the study design. One of such aspects is model structure, including the inputs and the outputs. Most statistical methods use a multivariate structure that splits the time series into 24 hourly subseries and model them independently. However, neural networks are well suited for vectorized outputs, so -- in theory -- a multi-output structure should be preferred. In particular, vectorized outputs should allow to significantly reduce computational times while increasing the performance level due to generalization properites \cite{lag:rid:sch:18}.

To tackle this issue, this study directly compares the two approaches using deep neural networks. As we show, besides the computational time benefit, the vectorized estimation also proves to be significantly better than the multi-model counterpart. For the sake of completion, the neural network results are compared with the LEAR model, a state-of-the-art parameter-rich linear structure with weights estimated using LASSO. The LEAR model is considered as it exhibits a good forecasting accuracy, it is much simpler to apply, and it is also widely used in the EPF literature \cite{uni:now:wer:16,zie:wer:18,Lago2020b}.

The other direction that the researchers choose is taking the well-known methods (both simpler, statistical and more complex ML ones) and improving them by various data preprocessing steps, such as variance stabilizing transformations \cite{dia:pla:16,uni:wer:zie:18} or separate modeling of the long-term seasonal component \cite{now:wer:16,lis:pel:18}. Another widely-adopted technique that improves the forecasts accuracy is forecast ensembling. The types of the forecasts that serve as the inputs are very diverse, ranging from e.g., multiple independent copies of the same model (with different different algorithm initialization point and local optimization) \cite{mar:uni:wer:19}, through the instances of the same model, but calibrated to different lenghts of historical data \cite{hub:mar:wer:19,mar:ser:wer:18}, to stacked models that use multiple forecasters and algorithm that chooses or aggregates the individual forecasts \cite{lag:rid:sch:18,naz:far:hei:sha:cat:18,zha:zha:li:tan:ji:19}.

The rest of the paper is structured as follows. In Section \ref{sec:Datasets} we present the datasets used in the study. In section \ref{sec:Methodology} we describe the methodology, with the emphasis on the Neural Network models and hyperparameter optimization. Section \ref{sec:Results} contains a detailed overview of obtained results, evaluated using two error measures: relative Mean Absolute Error (rMAE) and the discussion. In section \ref{sec:Conclusions} we conclude the results.

\section{Datasets}
\label{sec:Datasets}
The study uses the five open-access day-ahead market datasets proposed in \cite{Lago2020b}, which can be accessed using the \texttt{epftoolbox} \cite{benchmarkwebsite,epftoolboxdoc}, a \texttt{python} library for electricity price forecasting. Each dataset comprises three time series: one corresponding to the prices and two corresponding to exogenous series representing day-ahead forecasts (these depend on the dataset). The length of each datasets is equal to 2184 days (which translates to six years of 364 days -- or 52 weeks). All available time series are saved using the local time, and the daylight saving is treated appropriately, by either arithmetically averaging two values from the ``doubled'' hour or interpolating the neighboring values for the missing observation.

\begin{figure*}[tb]
	\centering
	\includegraphics[width=.9\textwidth]{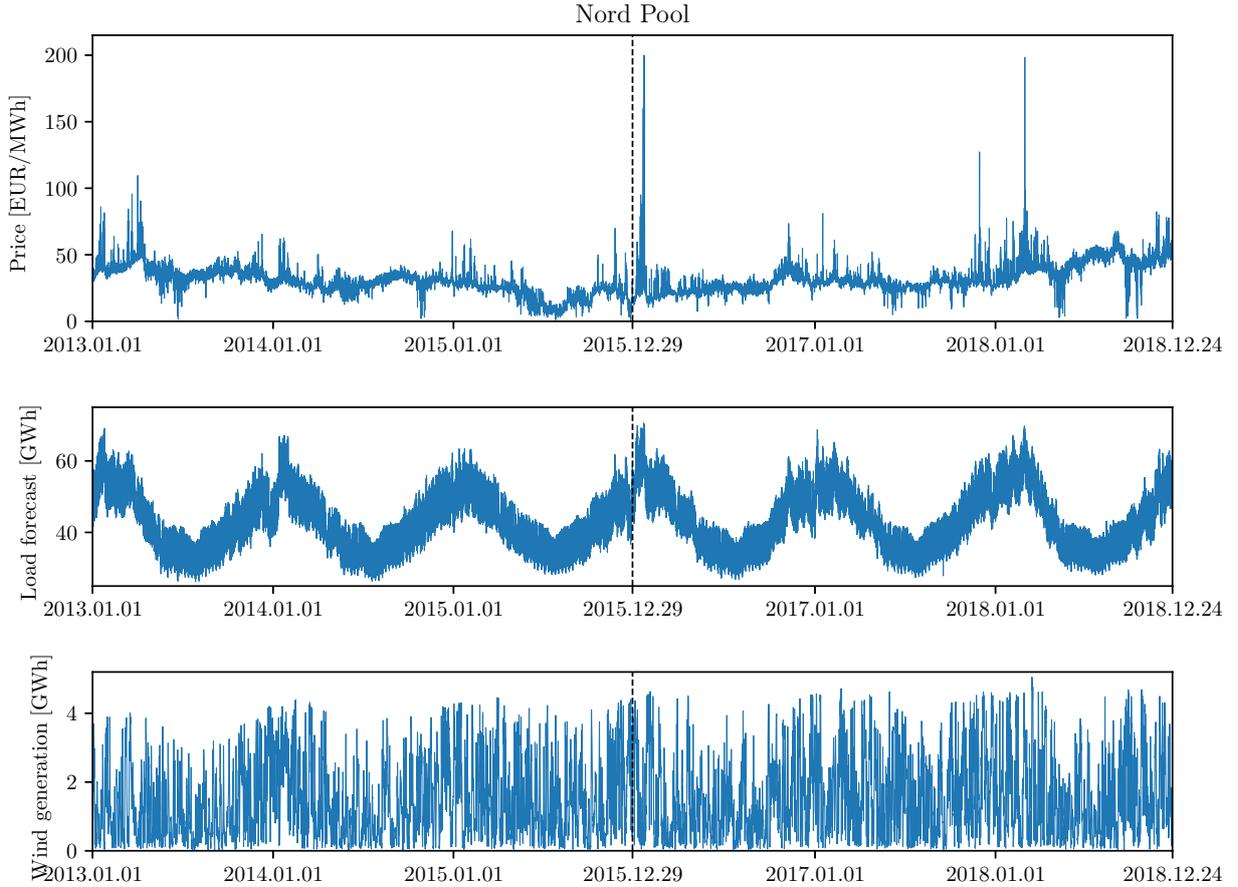}
	\caption{Nord Pool data.}
	\label{fig:dataNP}
\end{figure*}

The first dataset comes from the major European power market -- Nord Pool (NP) -- and spans the period from 01.01.2013 to 24.12.2018, see Fig. \ref{fig:dataNP}. The dataset contains hourly observations of system prices, day-ahead load forecast and day-ahead wind generation forecast. The yearly load pattern exhibits a strong high season during the winter months, and low season during summers, whereas wind generation presents more uniform behavior throughout the year. The price series shows several positive spikes, most apparent in more recent data.

The second dataset origins in the United States, specifically from the PJM (\emph{Pennsylvania-New Jersey-Maryland}) market. It covers the same time interval as Nord Pool -- from 01.01.2013 to 24.12.2018. The three time series are: the COMED (\emph{Commonwealth Edison} -- a zone located in the state of Illinois) zonal prices and two day-ahead load forecast series, one describing the system load, and the second one the COMED zone load. The price series shows a decrease in the number and magnitude of spikes in recent years. The exogenous series contain two high seasons during the year -- one in the winter and one in the summer.

The third dataset contains the French electricity prices from the EPEX SPOT market, the day-ahead load forecast in France and the day-ahead generation forecast in France. The dataset spans from 09.01.2011 to 31.12.2016. 

The fourth dataset is the EPEX SPOT in Belgium, and spans the same period as the French dataset. The exogenous series for this dataset, however, are taken from France instead (i.e. the exogenous series are exactly the same as for the French dataset) as the explanatory power of French features is stronger than that of Belgian features \cite{Lago2018}.

The last dataset describes the EPEX SPOT in Germany, and spans the period from 9th January 2012 to 31st December 2017. The two exogenous series comprise the zonal load forecast (Ampirion zone) and the wind generation forecast for the whole country, calculated as a sum of the off-shore and on-shore wind generation forecasts.

\begin{table}[th]
    \centering
    \caption{Periods of the data used for the out-of-sample test for each dataset.}
    \begin{tabular}{c|cc}
         \textbf{dataset} & \textbf{test period start} & \textbf{test period end} \\
         \hline
         NP & 27.12.2016 & 24.12.2018 \\
         PJM & 27.12.2016 & 24.12.2018 \\
         BE & 04.01.2015 & 31.12.2016 \\
         FR & 04.01.2015 & 31.12.2016 \\
         DE & 04.01.2015 & 31.12.2016
    \end{tabular}
    \label{tab:testperiods}
\end{table}

\section{Methodology}
\label{sec:Methodology}
\subsection{Forecast evaluation}
\label{ssec:evaluation}
The forecasts are evaluated using the relative Mean Absolute Error (rMAE), as suggested by \cite{Lago2020b}. The baseline forecast considered for the rMAE  is the seasonal na\"ive forecast, which sets the price for hour $h$ of day $d$ equal to the price corresponding to the same hour a week earlier, i.e., $\hat{p}_{d,h}^{\mathrm{naive}}=p_{d-7,h}$. The rMAE is given via the equation:
\begin{equation}
	\mathrm{rMAE} = \frac{\displaystyle\frac{1}{24N_\mathrm{d}}\sum_{d=1}^{N_\mathrm{d}}\sum_{h=1}^{24}|p_{d,h}-\hat{p}_{d,h}|}{\displaystyle\frac{1}{ 24N_\mathrm{d}}\sum_{d=1}^{N_\mathrm{d}}\sum_{h=1}^{24} |p_{d,h} - \hat{p}^{\mathrm{naive}}_{d,h} |}, \label{eqn:rmaeepf}
\end{equation}
where $N_d$ denotes the length of the test period (in days).
% where $e_j$ corresponds to the forecast error for time $j$, $j=T+1, T+2, \ldots, T+J$ are indices of the test hours, $T$ is the length of the training sample and $\widehat{Y}_t$ is a simple forecast used to scale the error. In this case, the scaling factor is the Na\"ve forecast \cite{nog:con:con:esp:02,wer:14} -- a similar-day method defined as follows. A Monday has identical prices as Monday of the previous week, similarly for Saturdays and Sundays. Tuesday is similar to the preceeding Monday, and the same rule applies for Wednesdays, Thursdays and Fridays.

The errors reported for all forecasts correspond to the out-of-sample test errors of a forecast done using a rolling calibration window scheme. The length of the out-of-sample test period is the last 728 days of each dataset (the specific dates are listed in Table \ref{tab:testperiods}. The models are calibrated using a 1456-days long (4 years $\times$ 364 days) rolling calibration window. 
% Note, that the evaluation is performed on the final outcomes, i.e., the forecasts that are transformed back, see Section \ref{ssec:VST}.

\subsubsection{Statistical significance of the differences}
\label{sssec:CPA}
Aside from the rMAE metric, the forecast accuracy is also verified using the Giacomini-White (GW) test for conditional predictive ability (CPA) \cite{gia:whi:06}, which can be seen as a generalization of the Diebold-Mariano test \cite{die:mar:95} which is widely used in the EPF literature \cite{uni:wer:zie:18,uni:now:wer:16,zie:wer:18}.

The GW test is a pairwise test that compares the outcomes of the models. In particular, having a pair of forecast series obtained using models $A$ and $B$, the test computes the multivariate loss series as:
\begin{equation*}
    \Delta^{\mathrm{A, B}}_{d} = ||\varepsilon^\mathrm{A}_d||_p - ||\varepsilon^\mathrm{B}_d||_p,
\end{equation*}
where $\varepsilon^\mathrm{Z}_d$ is the 24-dimensional vector of prediction errors of model Z for day $d$,  $||\varepsilon^\mathrm{Z}_d||_p = (\sum_{h=1}^{24} |\varepsilon^\mathrm{Z}_{d,h}|^p)^{1/p}$ is the $p$-th norm of that vector. Next, the \emph{p-value} of the GW test with null $H_0 : \phi = 0$ is computed for the regression:
\begin{equation}\label{eqn:GW}
    \Delta^{\mathrm{A,B}}_d=\boldsymbol{\phi}' {X}_{d-1} + \epsilon_d,
\end{equation}
where $X_{d-1}$ contains the information set on $d-1$, i.e., a constant and lags of $\Delta^{\mathrm{A,B}}_d$. In this study, for the sake of simplicty, all the tests are performed using $p=1$.

\subsection{Deep neural network models}
As described in the introduction, the goal of the study is to analyze the two main structures of DNNs that are typically used in the literature of EPF: single-output DNNs vs.\ DNNs with vectorized outputs.

\subsubsection{Input features}
\label{sec:inputdnn}
Before describing each model, let us define the input features that are considered. Independently of the model, the available input features to forecast the 24 day-ahead prices of day $d$, i.e.\ $\vc{p}_{d} = [p_{d,1},\ldots,p_{d,24}]^\top$, are the same:

\begin{itemize}
	\item Historical day-ahead prices of the previous three days and one week ago, i.e.\ $\vc{p}_{d-1}$, $\vc{p}_{d-2}$, $\vc{p}_{d-3}$, $\vc{p}_{d-7}$.
	\item The  day-ahead forecasts of the two variables of interest (see Section \ref{sec:Datasets} for details) for day $d$ available on day $d-1$, i.e.\ $\vc{{x}}^1_d=[x^1_{d,1},\ldots,x^1_{d,24}]^\top$ and $\vc{{x}}^2_d=[x^2_{d,1},\ldots,x^2_{d,24}]^\top$; note that the variables of interest are different for each market.
	\item Historical day-ahead forecasts of the variables of interest 
	the previous day and one week ago, i.e.\ $\vc{{x}}^1_{d-1}$, $\vc{{x}}^1_{d-7}$, $\vc{{x}}^2_{d-1}$, $\vc{{x}}^2_{d-7}$.
	\item A dummy variable $\vc{z}_d$ that represents the day of the week. This is with a multi-value input $z_d\in\{1,\ldots,7\}$. 
\end{itemize}

In total, we consider a total of 241 input features for each DNN model. As suggested in the literature \cite{Lago2018,Lago2020b}, the input features are optimized together with the hyperparameters using the tree Parzen estimator \cite{Bergstra2011}.

\subsubsection{DNN$_{24}$}

\begin{figure}[tb]
	\tikzset{%
		input neuron/.style={
			circle,
			draw,
			%     fill=blue!50,
			minimum size=.5cm
		},
		every neuron/.style={
		circle,
		draw,
		%     fill=blue!50,
		minimum size=.6cm
		},
		neuron missing/.style={
			draw=none, 
			scale=1.2,
			text height=.25cm,
			execute at begin node=\color{black}$\vdots$
		},
		sigmoid/.style={path picture= {
				\begin{scope}[x=.7pt,y=7pt]
					\draw plot[domain=-6:6] (\x,{1/(1 + exp(-\x))-0.5});
				\end{scope}
			}
		},
		linear/.style={path picture= {
				\begin{scope}[x=5pt,y=5pt]
					\draw plot[domain=-1:1] (\x,\x);
				\end{scope}
			}
		},
	}

	\centering
	\begin{tikzpicture}[x=1cm, y=.65cm, >=stealth, scale=.9]
	
	% \foreach \m/\l [count=\y] in {1,2,3,4,5,6,7,8}
	% \node [input neuron/.try, neuron \m/.try] (input-\m) at (0,2.5-\y) {\tiny24};
	% \foreach \m/\l [count=\y] in {1,2,3,4,5,6,7,8}
	\node [input neuron/.try, neuron 1/.try] (input-1) at (-1,2.5-1) {\makebox[7pt]{}};
	\node [input neuron/.try, neuron 2/.try] (input-2) at (-1,2.5-2) {\makebox[7pt]{}};
	\node [input neuron/.try, neuron 3/.try] (input-3) at (-1,2.5-3) {\makebox[7pt]{}};
	\node [input neuron/.try, neuron 4/.try] (input-4) at (-1,2.5-4) {\makebox[7pt]{}};
	\node [input neuron/.try, neuron 5/.try] (input-5) at (-1,2.5-5) {\makebox[7pt]{}};
        % \node [input neuron/.try, neuron 6/.try] (input-6) at (-1,2.5-6) {\makebox[8.5pt]{}};
	\node [input neuron/.try, neuron 6/.try] (input-6) at (-1,2.5-7) {\makebox[7pt]{}};
	\node [input neuron/.try, neuron 7/.try] (input-7) at (-1,2.5-8) {\makebox[7pt]{}};
	
	\node [] at (-1,2.5-6) {\ldots};

	\foreach \m [count=\y] in {1,2,3,4,5}
	\node [every neuron/.try, neuron \m/.try, fill= black!20, sigmoid ] (hidden-\m) at (2,2.5-\y*1.5) {};
	
	\foreach \m [count=\y] in {1,2,3}
	\node [every neuron/.try, neuron \m/.try, fill= black!20, sigmoid ] (hidden2-\m) at (4,2-\y*2) {};
	
	\foreach \m [count=\y] in {1}
	\node [every neuron/.try, neuron \m/.try, fill= blue!20, linear ] (output-\m) at (6,-1.0-\y) {};
	
	% \foreach \l [count=\i] in {{$\mathbf{X}_{d-1}$},{$\mathbf{X}_{d-2}$},{$\mathbf{X}_{d-7}$},$X_{d-1}^{\min}$,{$\mathbf{Z}_{d}$}, $D_1$, $D_2$}
		\foreach \l [count=\i] in {$X_1$,$X_2$,$X_3$,$X_4$,$X_5$,$X_{n-1}$,$X_n$}
	\draw [<-] (input-\i) -- ++(-1.2,0)
	node [above, midway] {\footnotesize\l};
	
	\foreach \l [count=\i] in {1,2,3,4,5}
	\node [above] at (hidden-\i.north) {};
	
	\foreach \l [count=\i] in {1}
	\draw [->] (output-\i) -- ++(1.2,0)
	node [above, midway] {\footnotesize{$\hat p_{d, h_0}$}};
	
	\foreach \i in {1,...,7}
	\foreach \j in {1,...,5}
	\draw [->] (input-\i) -- (hidden-\j);
	
	\foreach \i in {1,...,5}
	\foreach \j in {1,...,3}
	\draw [->] (hidden-\i) -- (hidden2-\j);
	
	\foreach \i in {1,...,3}
	\foreach \j in {1}
	\draw [->] (hidden2-\i) -- (output-\j);
	
	\foreach \l [count=\x from 1] in {Hidden, Hidden, Output}
	\node [align=center, above] at (\x*2,2) {\footnotesize\l \\[-3pt] \footnotesize{layer}};
	
	\node [align=center, above] at (-1,2) {\footnotesize Input \\[-3pt] \footnotesize{layer}};
	
	\end{tikzpicture}	
	\caption{\label{fig:dnn24}Schematic representation of the single-output DNN. Note, that the input structure depends on the feature selection performed during the hyperparameter optimization and may differ for different hour-specific models.}
\end{figure}
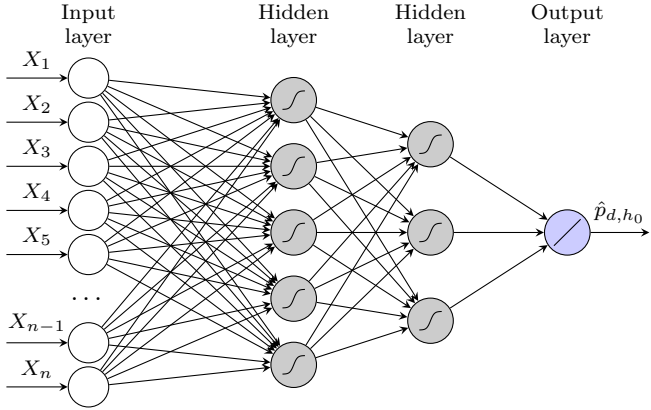

The first DNN model represents the open-source DNN proposed in \cite{Lago2020b} and it is denoted by DNN$_{24}$ as it uses an explicit vectorized output to model the 24 hours of the day. Its main advantage is that it requires less computational resources and it is  less prone to overfit as it can generalize better. As in \cite{Lago2020b}, the features and hyperparameters of the model are optimized together using the tree Parzen estimator \cite{Bergstra2011}. A representation of this model is provided in Figure \ref{fig:dnn24}.

% For the single-output structure, the hyperparameter optimization is done separately for each of the 24 models. The hyperparameter optimization procedure is fully automated, and is based on a parameter-rich structure, with explanatory variables identical to those of the LASSO benchmark (see Eq. (\ref{eqn:LASSO})), see Section \ref{ssec:hyperparameters} for details.

\subsubsection{DNN$_{1}$}
\begin{figure}[tb]
	\tikzset{%
		input neuron/.style={
			circle,
			draw,
			%     fill=blue!50,
			minimum size=.5cm
		},
		every neuron/.style={
		circle,
		draw,
		%     fill=blue!50,
		minimum size=.6cm
		},
		neuron missing/.style={
			draw=none, 
			scale=1.2,
			text height=.25cm,
			execute at begin node=\color{black}$\vdots$
		},
		sigmoid/.style={path picture= {
				\begin{scope}[x=.7pt,y=7pt]
					\draw plot[domain=-6:6] (\x,{1/(1 + exp(-\x))-0.5});
				\end{scope}
			}
		},
		linear/.style={path picture= {
				\begin{scope}[x=5pt,y=5pt]
					\draw plot[domain=-1:1] (\x,\x);
				\end{scope}
			}
		},
	}

	\centering
	\begin{tikzpicture}[x=1cm, y=.65cm, >=stealth, scale=.9]
	
	% \foreach \m/\l [count=\y] in {1,2,3,4,5,6,7,8}
	% \node [input neuron/.try, neuron \m/.try] (input-\m) at (0,2.5-\y) {\tiny24};
	% \foreach \m/\l [count=\y] in {1,2,3,4,5,6,7,8}
	\node [input neuron/.try, neuron 1/.try] (input-1) at (-1,2.5-1) {\makebox[7pt]{}};
	\node [input neuron/.try, neuron 2/.try] (input-2) at (-1,2.5-2) {\makebox[7pt]{}};
	\node [input neuron/.try, neuron 3/.try] (input-3) at (-1,2.5-3) {\makebox[7pt]{}};
	\node [input neuron/.try, neuron 4/.try] (input-4) at (-1,2.5-4) {\makebox[7pt]{}};
	\node [input neuron/.try, neuron 5/.try] (input-5) at (-1,2.5-5) {\makebox[7pt]{}};
        % \node [input neuron/.try, neuron 6/.try] (input-6) at (-1,2.5-6) {\makebox[8.5pt]{}};
	\node [input neuron/.try, neuron 6/.try] (input-6) at (-1,2.5-7) {\makebox[7pt]{}};
	\node [input neuron/.try, neuron 7/.try] (input-7) at (-1,2.5-8) {\makebox[7pt]{}};
	
	\node [] at (-1,2.5-6) {\ldots};

	\foreach \m [count=\y] in {1,2,3,4,5}
	\node [every neuron/.try, neuron \m/.try, fill= black!20, sigmoid ] (hidden-\m) at (2,2.5-\y*1.5) {};
	
	\foreach \m [count=\y] in {1,2,3}
	\node [every neuron/.try, neuron \m/.try, fill= black!20, sigmoid ] (hidden2-\m) at (4,2-\y*2) {};
	
	\foreach \m [count=\y] in {1,2,3,5}
	\node [every neuron/.try, neuron \m/.try, fill= blue!20, linear ] (output-\y) at (6,2.5-\m*1.5) {};
	
	% \foreach \l [count=\i] in {{$\mathbf{X}_{d-1}$},{$\mathbf{X}_{d-2}$},{$\mathbf{X}_{d-7}$},$X_{d-1}^{\min}$,{$\mathbf{Z}_{d}$}, $D_1$, $D_2$}
	\foreach \l [count=\i] in {$X_1$,$X_2$,$X_3$,$X_4$,$X_5$,$X_{n-1}$,$X_{n}$}
	\draw [<-] (input-\i) -- ++(-1.2,0)
	node [above, midway] {\footnotesize\l};
	
	\node [] at (6,-3.5) {\ldots};
	
	\foreach \l [count=\i] in {1,2,3,4,5}
	\node [above] at (hidden-\i.north) {};
	
	\foreach \l [count=\i] in {1,2,3,24}
	\draw [->] (output-\i) -- ++(1.2,0)
	node [above, midway] {\footnotesize{$\hat p_{d, \l}$}};
	
	\foreach \i in {1,...,7}
	\foreach \j in {1,...,5}
	\draw [->] (input-\i) -- (hidden-\j);
	
	\foreach \i in {1,...,5}
	\foreach \j in {1,...,3}
	\draw [->] (hidden-\i) -- (hidden2-\j);
	
	\foreach \i in {1,...,3}
	\foreach \j in {1,2,3,4}
	\draw [->] (hidden2-\i) -- (output-\j);
	
	\foreach \l [count=\x from 1] in {Hidden, Hidden, Output}
	\node [align=center, above] at (\x*2,2) {\footnotesize\l \\[-3pt] \footnotesize{layer}};
	
	\node [align=center, above] at (-1,2) {\footnotesize Input \\[-3pt] \footnotesize{layer}};
	
	\end{tikzpicture}	
	\caption{\label{fig:dnn1}Deep multi-output Nerual Network schematics, with sigmoid activation in the hidden layer. Note, that the input structure depends on the feature selection performed during the hyperparameter optimization.}
\end{figure}
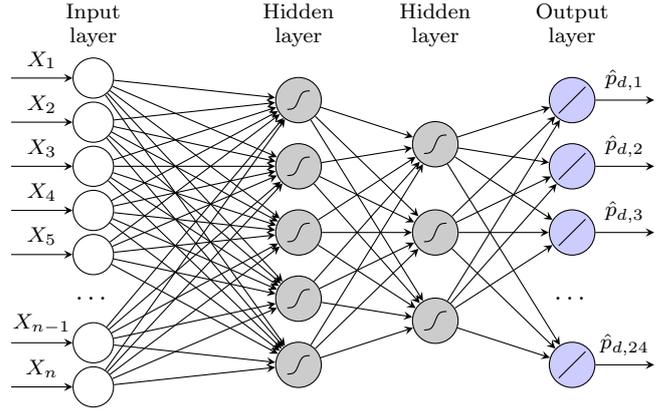

The second approach, denoted here as DNN$_1$, considers 24 independent separate models (one for each hour of the day). Such an approach is much slower (roughly 20x) than DNN$_{24}$; however, it has the advantage of being able to select different hyperparameters/input features for each specific hours. It is important to note that this ability to specialize, while useful with parsimonious linear models, it might not be ideal for DNNs where generalization is a more desirable property as it prevents overfitting. A representation of this model is provided in Figure \ref{fig:dnn1}.

 \subsubsection{Implementation details}
Both models are implemented in Keras \cite{keras}, with the estimation performed using the Adam optimization algorithm \cite{kin:ba:14} and early stopping to prevent overfitting. In both cases, following standard practice in the literature of EPF \cite{lag:rid:sch:18,Lago2020b,Lago2018}, the depth of the networks is fixed to two hidden layers. In addition, both models have an optional dropout hyperparameter that selects the dropout rate.

As in \cite{Lago2020b}, both DNNs are estimated using 4 year long (208 weeks) rolling calibration window, with 42  weeks randomly selected as a validation set used for early stopping, and the remaining 166 weeks used for the training.

\subsubsection{Hyperparameter optimization}
\label{ssec:hyperparameters}
The DNNs require a hyperparameter optimization procedure \cite{Lago2020b}. The methodology is performed once, using the data from before the first day of the out-of-sample test window, and the resulting hyperparameters are used throughout the whole testing period. As suggested in the literature \cite{Lago2018,Lago2020b}, this step is performed using the tree Parzen estimator \cite{Bergstra2011}.

 Out of a 208 weeks available for model determination and validation, the first 166 weeks are used as a training set, and the remaining 42 weeks -- as a validation set. Unlike for the final evaluation, the 42 weeks  of the validation dataset are not randomly selected but the validation to select the hyperparameters is performed on the most recent portion of the data (we refer to \cite{Lago2020b} for details and motivation behind this procedure). As indicated in Section \ref{sec:inputdnn}, the input features are optimized together with the hyperparameters.

The hyperparameters that are subject to optimization include: 1) the number of neurons of each hidden layer, 2) the activation function used in the hidden layers, 3) the dropout rate, 4) the learning rate, 5) the use of batch normalization, 6) the data preprocessing technique, 7) the type of network weights initialization and 8) the coefficient for kernel regularization of layers. 

To select which of the 241 available input features are relevant, the method models the input features as hyperparameters \cite{Lago2018} and employs 11 decision variables, i.e.\ 11 hyperparameters:
\begin{itemize}
	\item Four binary hyperparameters (1-4) that indicate whether or not to include the historical day ahead prices $\vc{p}_{d-1}$, $\vc{p}_{d-2}$, $\vc{p}_{d-3}$, $\vc{p}_{d-7}$.  The selection is done per day\footnote{This is done for the sake of simplicity to speed up the optimization procedure of the feature selection. In particular, an alternative could be to use a binary hyperparameter for each individual historical prices; however, is most markets, that would mean using 24 as many hyperparameters as there are 24 different prices per day.}, e.g.\ the algorithm either selects all the prices $\vc{p}_{d-j}$ of $j$ days ago or it cannot select any price from day $d-j$, hence the four hyperparameters.
	\item Two binary hyperparameters (5-6) that indicate  whether or not to include each of the day-ahead forecasts $\vc{x}^1_{d}$ and $\vc{x}^2_{d}$. As with the past prices, this is done for the whole day, i.e.\ a hyperparameter either selects all the elements in $\vc{x}^j_{d}$ or none.
	\item Four binary hyperparameters (7-10) that indicate whether or not to include the historical day-ahead forecasts $\vc{x}^1_{d-1}$, $\vc{x}^2_{d-1}$, $\vc{x}^1_{d-7}$, and $\vc{x}^2_{d-7}$. This selection is also done per day.
	\item One binary hyperparameter (11) that indicates whether or not to include the variable $z_d$ representing the day of the week.
\end{itemize}

It is important to note, that for the single-output networks, the procedure is performed independently for all 24 hours of the day, resulting in 24 sets of input variables and network hyperparameters. 

\subsubsection{Data preprocessing for neural network models}
One of the DNNs hyperparameters is the data optimization technique used. There are six choices:
\begin{itemize}
	\item no normalization -- raw data is passed to the network,
	\item linear mapping of each feature to $[0, 1]$,
	\item linear mapping of each feature to $[-1, 1]$,
	\item normalization using mean and standard deviation,
	\item normalization using median and MAD (see Eqn. (\ref{eqn:mediannorm}))
	\item normalization using median and MAD, with asinh VST applied.
\end{itemize}
The method is chosen independently for the inputs and training targets. Each input is normalized separately from other inputs, but using the same method.

\subsection{Benchmark model}
\label{ssec:LASSO}
The model that sets down the baseline scores is a parameter-rich linear model estimated using the least absolute shrinkage and selection operator (LASSO) \cite{tib:96}. After \cite{Lago2020b}, we refer to it as LEAR model (abbreviation of LASSO estimated autoregressive). 

\subsubsection{Input features}
The LEAR model uses the exactly same inputs as the DNN models. However, there is a difference in the weekday dummies. DNN models comprise only one dummy variable that takes values $0,\,1,\,\ldots,6$, respectively for Monday, Tuesday, \ldots, Sunday. The LEAR model usings a binary vector of 7 values that sets to 1 the respective day of the wek.

\subsubsection{Equation}
The benchmark model uses a multivariate framework -- i.e., the implicit $day\times hour$ structure where 24 models are trained (one for each hour of the day) \cite{zie:wer:18}. However, using the parameter-rich structure, the right-hand side of the equation is identical for each hour, allowing the model to consider intraday price dynamics. The preprocessed (see Section \ref{ssec:VST}) price $p$ for hour $h_0$, day $d$ is modeled as follows.
\begin{align}
	\label{eqn:LASSO}
	p_{d,h_0} &= \sum_{h=1}^{24}\beta_{h,h_0}p_{d-1, h} + \sum_{h=1}^{24}\beta_{24 + h,h_0}p_{d-2, h}+\nonumber\\
		  &+ \sum_{h=1}^{24}\beta_{48 + h,h_0}p_{d-3, h} + \sum_{h=1}^{24}\beta_{72 + h,h_0}p_{d-7, h} + \nonumber\\
		  &+ \sum_{h=1}^{24}\beta_{96 + h,h_0}z_{d, h}^{(1)} + \sum_{h=1}^{24}\beta_{120 + h,h_0}z_{d-1, h}^{(1)}+\nonumber\\
		  &+ \sum_{h=1}^{24}\beta_{144 + h,h_0}z_{d-7, h}^{(1)}+ \sum_{h=1}^{24}\beta_{168 + h,h_0}z_{d, h}^{(2)} +\nonumber\\
		  &+ \sum_{h=1}^{24}\beta_{192 + h,h_0}z_{d-1, h}^{(2)}+\sum_{h=1}^{24}\beta_{216 + h,h_0}z_{d-7, h}^{(2)} +\nonumber\\
		  &+  \sum_{i=1}^7\beta_{240+i,h_0}D_i + \varepsilon_{d,h_0},
\end{align}
where $z_{d,h}^{(i)}$ is the preprocessed value of $i$-th exogenous series for day $d$, hour $h$, and $D_i$ is a weekday dummy (binary) variable that marks the $i$-th day of the week.

\subsubsection{Variance Stabilizing Transformation}
\label{ssec:VST}
The LEAR benchmark models were run on preprocessed data. The transformation used was \emph{area hyperbolic sine} {(asinh)}, as suggested in the review of various variance stabilizing transformations (VST) in EPF \cite{uni:wer:zie:18}. This allows the linear models to achieve higher forecasting performance, especially in the cases of price spikes.

The application of the VST is a two-step procedure, firstly, the data series are normalized, and secondly, the transformation function is applied. After the forecasting, the back transformation is perfomed in an inverse order, i.e., the outcome is first passed through an inverse transformation function appropriate for the VST chosen, and lastly the value is denormalized.

The normalization is performed for each input variable independently (i.e., the series of prices at hour 9 has different normalization factors than hour 10). The procedure for each input series $X_{d,h}$ consists of subtracting the sample median and dividing by the sample Median Absolute Deviation (MAD), corrected by the 75-th percentile $z_{0.75}$ of the standard normal distribution:
\begin{equation}
	\label{eqn:mediannorm}
	x_{d,h} = \frac{z_{0.75}}{\textrm{MAD}(X)} (X_{d,h} - \textrm{median}(X)),
\end{equation}
where $X$ without the subscript denotes the vector of observations in the dataset. After that, the asinh transformation function is applied:
\begin{equation}
	Y_{d,h} = \textrm{asinh}(x_{d,h}) = \log\left(x_{d,h} + \sqrt{x_{d,h}^2+1}\right).
\end{equation}
After the forecasting, the predictions are back transformed:
\begin{equation}
	\widehat{P}_{d,h} = \frac{\textrm{MAD(P)}}{z_{0.75}}\,\sinh(p_{d,h}) + \textrm{median}(P),
\end{equation}
where $p$ denotes the transformed response variable series, $P$ and $\widehat P$ -- a non-transformed response variable series, respectively the in-sample part and the final forecast. Please note that the dummy variables included in the model are not normalized.

\section{Results}
\label{sec:Results}
\begin{table}[tbh]
	\centering
	\caption{\label{tab:rmae}Table lists joint rMAE for all 24 hours of the day. Bold result correspond to the lowest error for a given dataset.}
	\begin{tabular}{c|ccc}
		dataset & LEAR & DNN$_{24}$ & DNN$_1$ \\
		\hline
		NP & 0.482 & \textbf{0.472} & \textbf{0.472} \\
		PJM & 0.492 & \textbf{0.477} & 0.495 \\
		BE & 0.655 & \textbf{0.611} & 0.649 \\
		FR & 0.596 & \textbf{0.575} & 0.628 \\
		DE & 0.452 & \textbf{0.447} & 0.516 \\
	\end{tabular}
\end{table}

\begin{figure*}[!ht]
	\begin{subfigure}[t]{\textwidth}
		\begin{center}
			\includegraphics[width=0.495\textwidth]{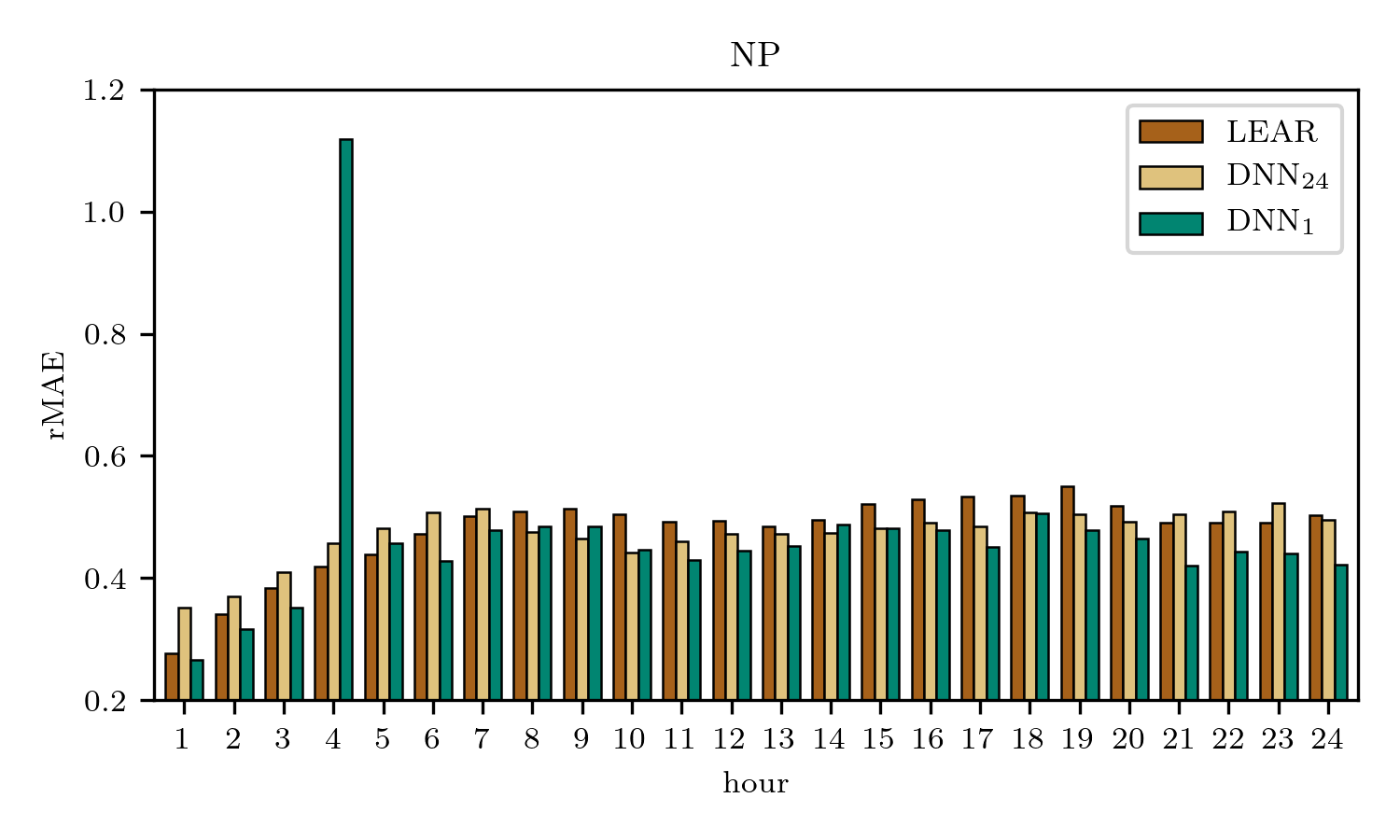}
% 			\caption{EPEX-FR}
			%\label{fig:data_fr}	
			\includegraphics[width=0.495\textwidth]{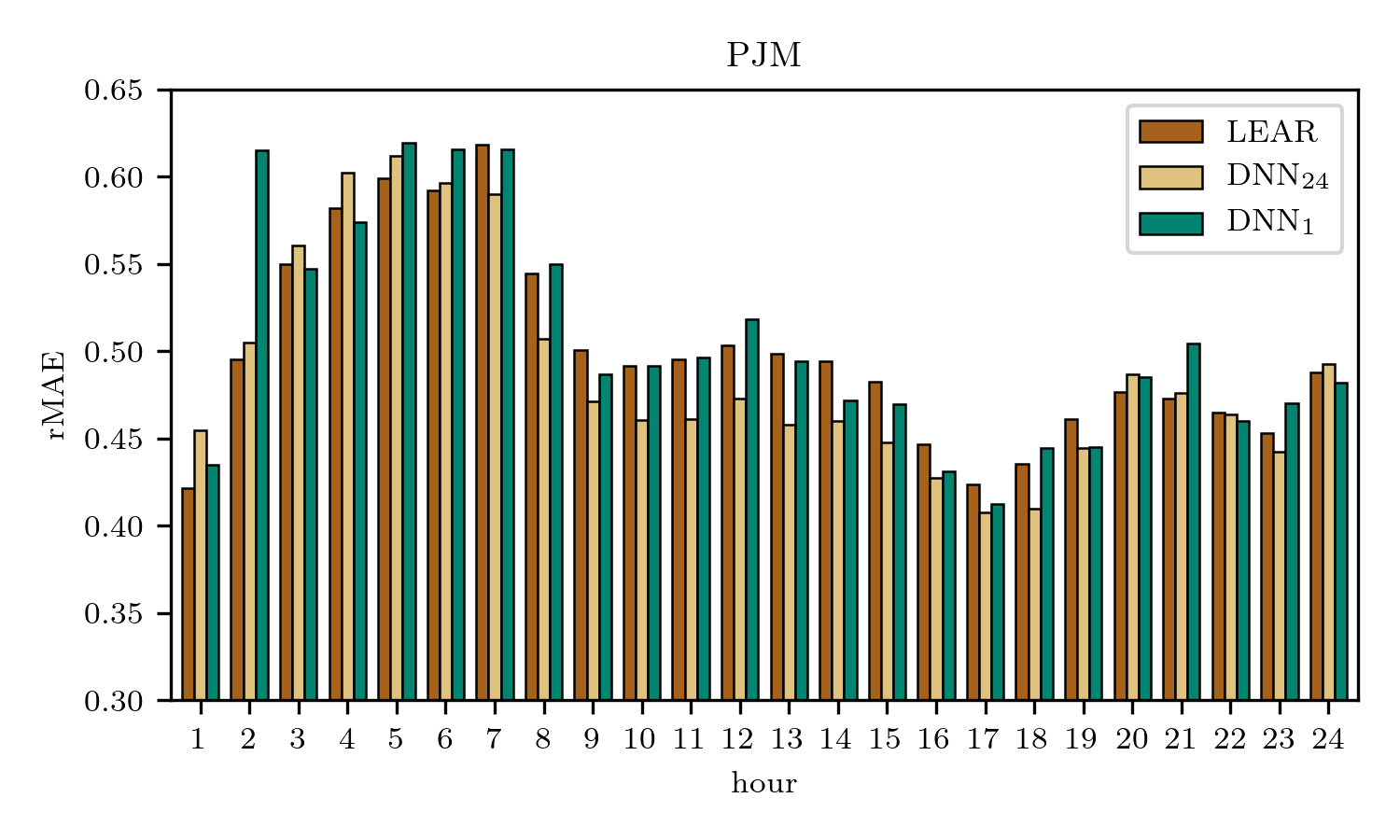}
% 			\caption{EPEX-FR}
			%\label{fig:data_de}			
		\end{center}
	\end{subfigure}
	\caption{Hourly rMAE errors for the Nord Pool and PJM markets.}
	\label{fig:hourlyPJMNP}
\end{figure*}
This section presents the results for the five distinct datasets. All reported metrics refer to the rMAE with the seasonal naive forecast as presented in (\ref{eqn:rmaeepf}) (see Section \ref{ssec:evaluation}). Table \ref{tab:rmae} presents the errors jointly for all 24 hours of the day, for all tested models and 5 test datasets. What can be immediately observed, is the significantly better performance of DNN$_{24}$ compared to the remaining two methods.

\begin{figure*}[!ht]
	\begin{subfigure}[t]{\textwidth}
		\begin{center}
			\includegraphics[width=0.495\textwidth]{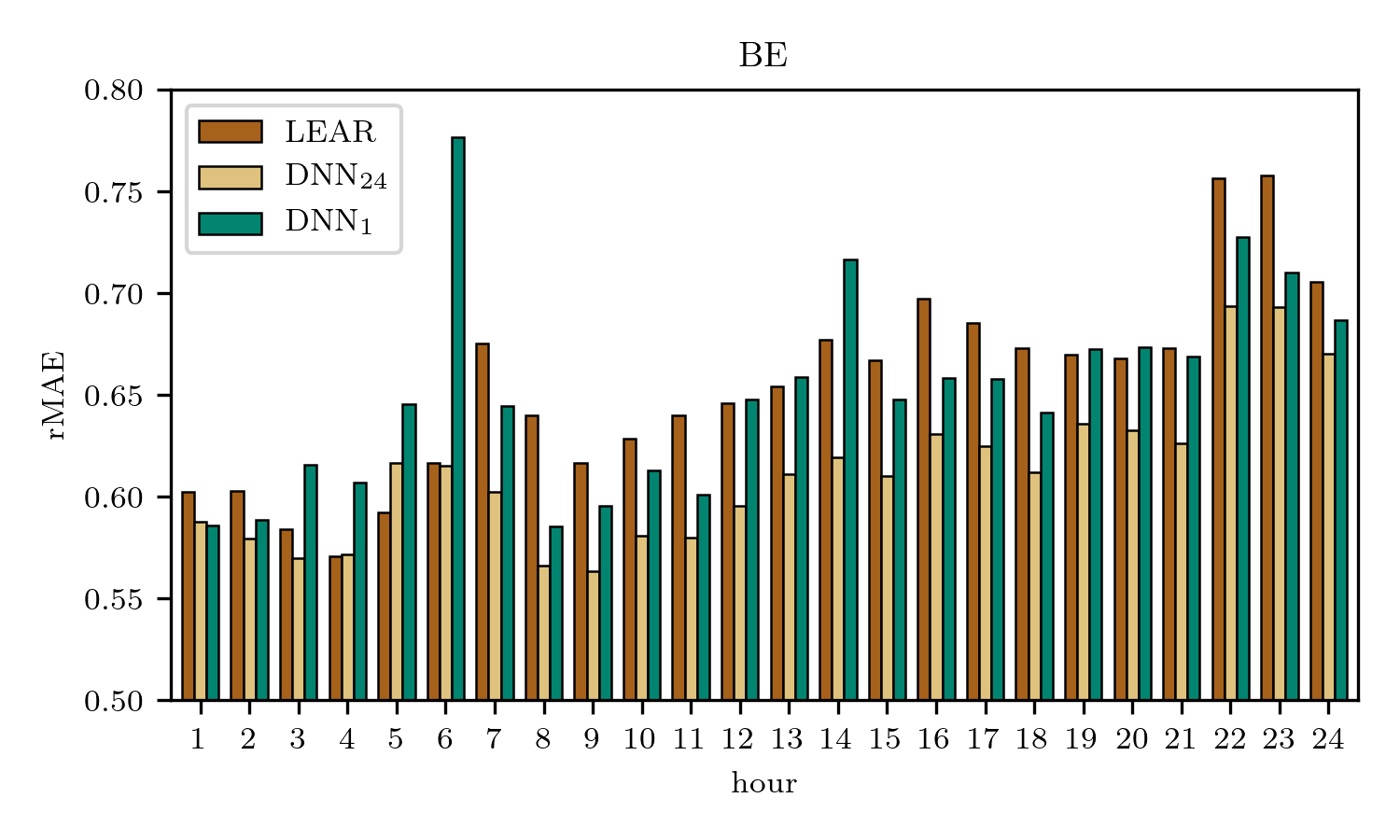}
% 			\caption{EPEX-FR}
			%\label{fig:data_fr}	
			\includegraphics[width=0.495\textwidth]{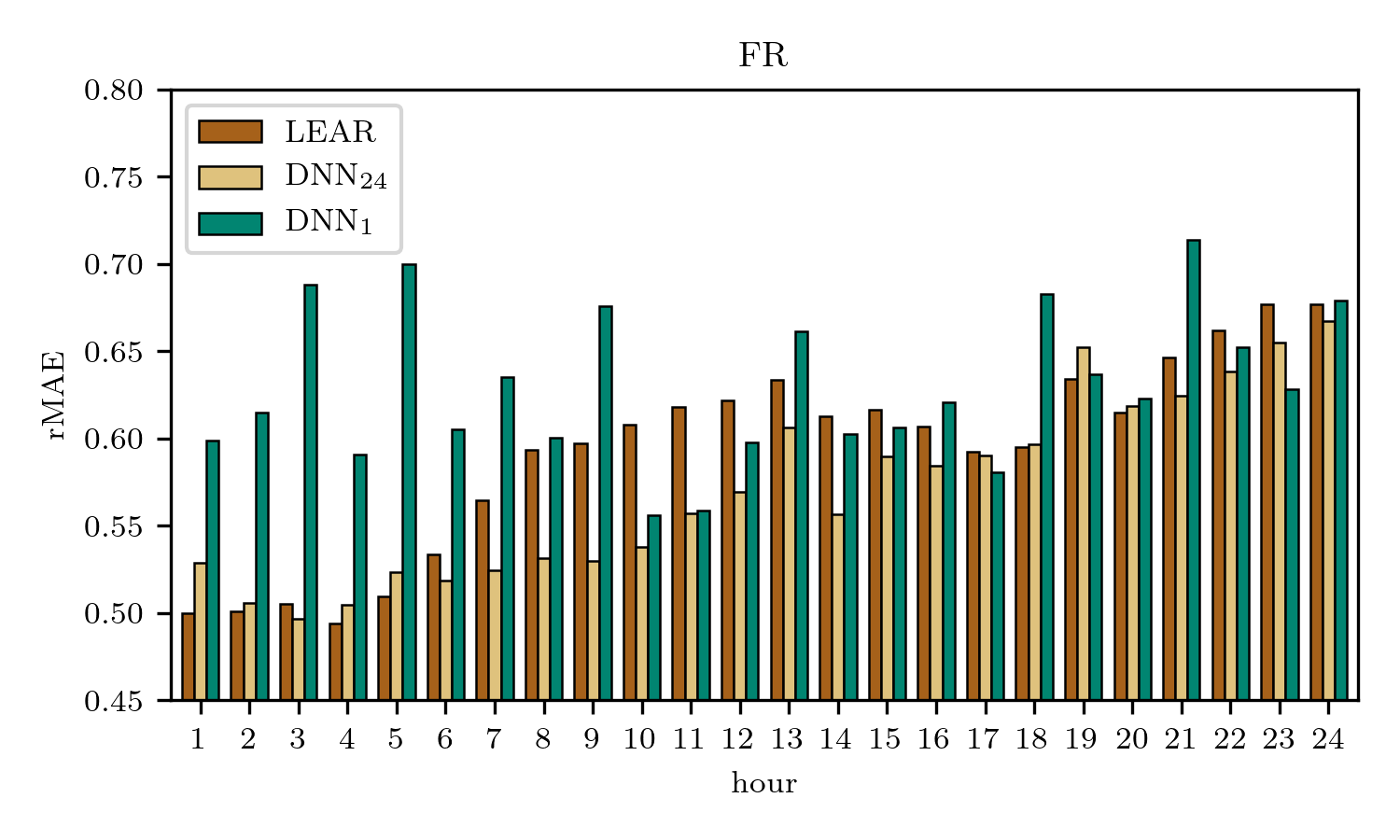}
% 			\caption{EPEX-FR}
			%\label{fig:data_de}			
		\end{center}
	\end{subfigure}
	\begin{subfigure}[t]{\textwidth}
		\begin{center}
			\includegraphics[width=0.495\textwidth]{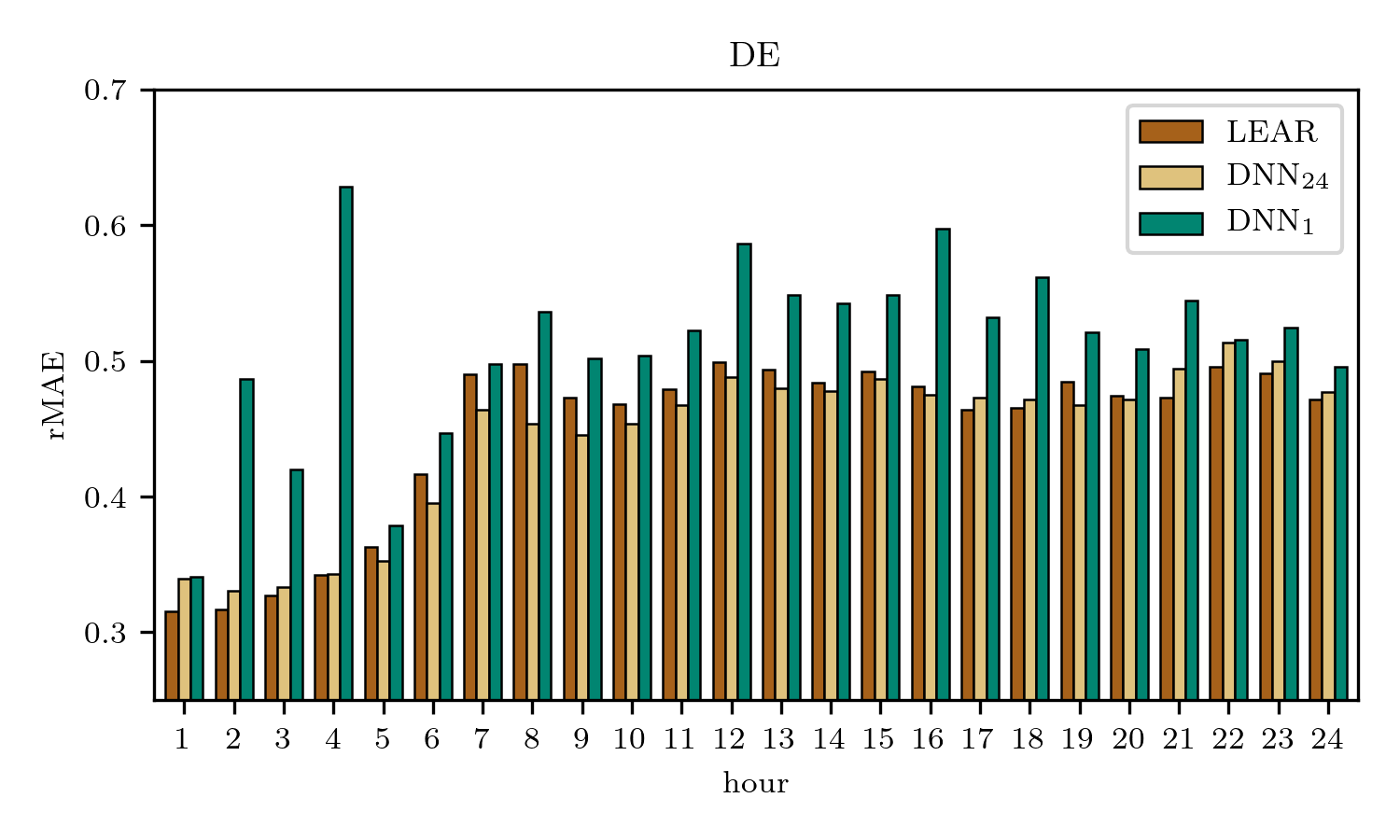}
% 			\caption{EPEX-FR}
			%\label{fig:data_de}			
		\end{center}
	\end{subfigure}	
	\caption{Hourly rMAE errors for EPEX-BE, EPEX-FR, and EPEX-DE markets.}
	\label{fig:hourlyBEFRDE}
\end{figure*}

Figures \ref{fig:hourlyPJMNP} and \ref{fig:hourlyBEFRDE} extend the analysis by providing the rMAE of each model for each hour of the day. An interesting observation is that the worse performance of DNN$_1$ model (relative to the DNN$_{24}$) is observed mostly at specific hours of the day.

\subsection{Single vs vectorized output}
The overall worse performance of the single-output approach relative to the multi-ouput counterpart is observed for almost all of the test datasets.
In particular, for the the EPEX-BE, EPEX-FR, and EPEX-DE datasets, the DNN$_1$ model exhibits an erroneous results for nearly all hours, typically during the early off-peak period, see Figure \ref{fig:hourlyBEFRDE}.

In the other two dataset, the difference in performance is lesser. Particularly, the underperfomance of the DNN$_1$ model is less pronounced on the PJM test set (Fig. \ref{fig:hourlyPJMNP} top), with only one hour  having a visibly higher error. In the Nord Pool dataset, both models score identically when all 24 hours of the day are considered. However, for hour 4, the single-output model had a predictive performance lower than the one of the naive benchmark (as indicated by rMAE greater than 1). For most of the other hours, the DNN$_1$ model outperforms the DNN$_{24}$ one.

Despite the early stopping criterion and dropout implemented to avoid overfiting, the large difference between DNN$_1$ and DNN$_{24}$ at specific hours of the day can be easily explained  due to the overfitting. In the multi-output setting, a part of network is shared between all of the 24 output nodes. This part of the network ideally models the factors that are similar for all hours and is able to generalize predictions without overfiting. In the single-output scenario, all the neurons are specialized to a single hour and the risk of overfiting is larger.

\subsection{Baseline model}

As a final comparison, it is important to analyze the performance of the considered models against the state-of-the-art baseline model. As suggested in \cite{Lago2020b}, such a comparison ensures that the evaluated models are sound and reasonable for EPF. 

When compared with the baseline model, the DNN$_{24}$ model outperforms the LEAR model for the morning peak hours on all 5 datasets and for the evening peak hours on PJM, Nord Pool and EPEX-BE. By contrast, while on average it is better, it is outperformed during the first couple hours of the day on four out of the five datasets.

The average performance of LEAR approach is comparable with the DNN$_1$ performance for Nord Pool, PJM and EPEX-BE data, and is closer to the better DNN$_{24}$ model for the EPEX-FR and EPEX-DE datasets.

%\gray{In short, as can be observed, from the results, both models provide similar or better results than the LEAR model and are thus reasonable models for EPF.} % It is important to note, that the over-fitting occurs in the DNN$_1$ model despite using several techniques that aim to prevent that, namely early stopping condition during training based on the model performance on the validation dataset and feature selection performed separately for each hour. On the other hand, the LARS benchmark did not exhibit such issues, showing that single-output parameter-rich structures with feature selection can perform well in the tested setting.

In short, as can be observed, from the results, both models provide similar or better results than the LEAR model. However, it is important to note, that the over-fitting occurs in the DNN$_1$ model despite using several techniques that aim to prevent that, namely early stopping condition during training based on the model performance on the validation dataset and feature selection performed separately for each hour. Conversely, the LEAR benchmark did not exhibit such issues, showing that single-output parameter-rich structures with feature selection can perform well in the tested setting -- however, not as well as the DNN$_{24}$.

\subsection{Test for Conditional Predictive Ability}

\begin{figure*}[ht]
\begin{subfigure}[t]{0.325\textwidth}
\begin{center}
	\includegraphics[width=\textwidth]{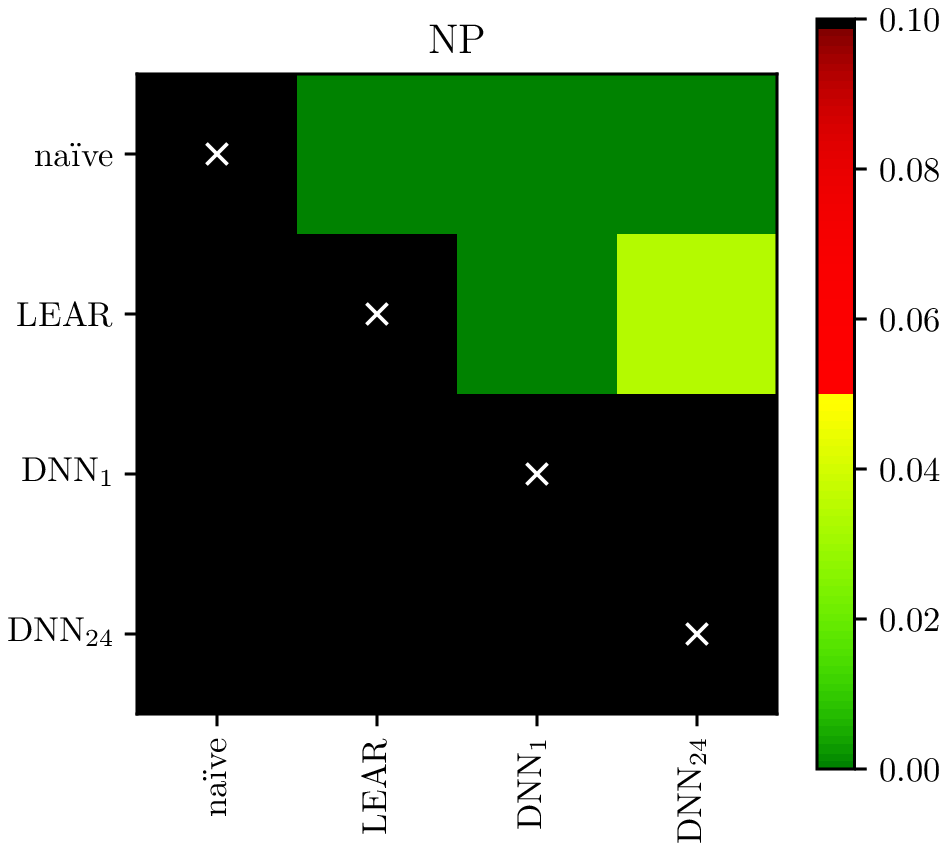}
% 	\caption{Nord pool}
\end{center}
\end{subfigure}
\begin{subfigure}[t]{0.325\textwidth}
\begin{center}
	\includegraphics[width=\textwidth]{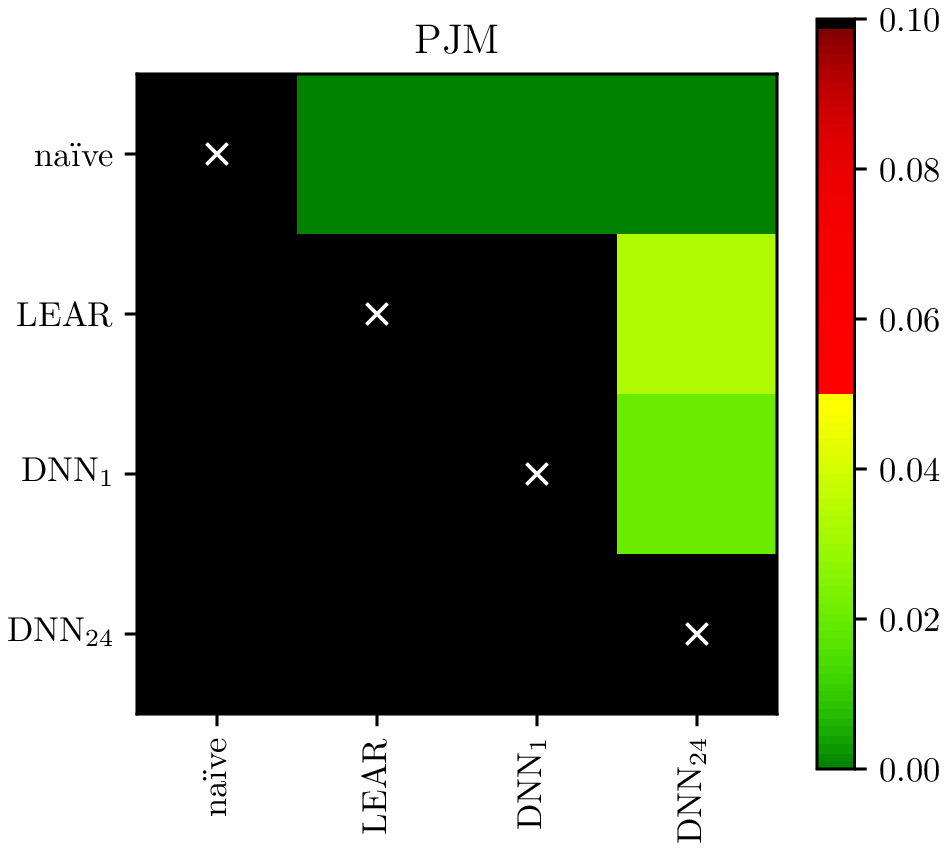}
% 	\caption{PJM}
\end{center}
\end{subfigure}
	\begin{subfigure}[t]{0.325\textwidth}
		\begin{center}
			\includegraphics[width=\textwidth]{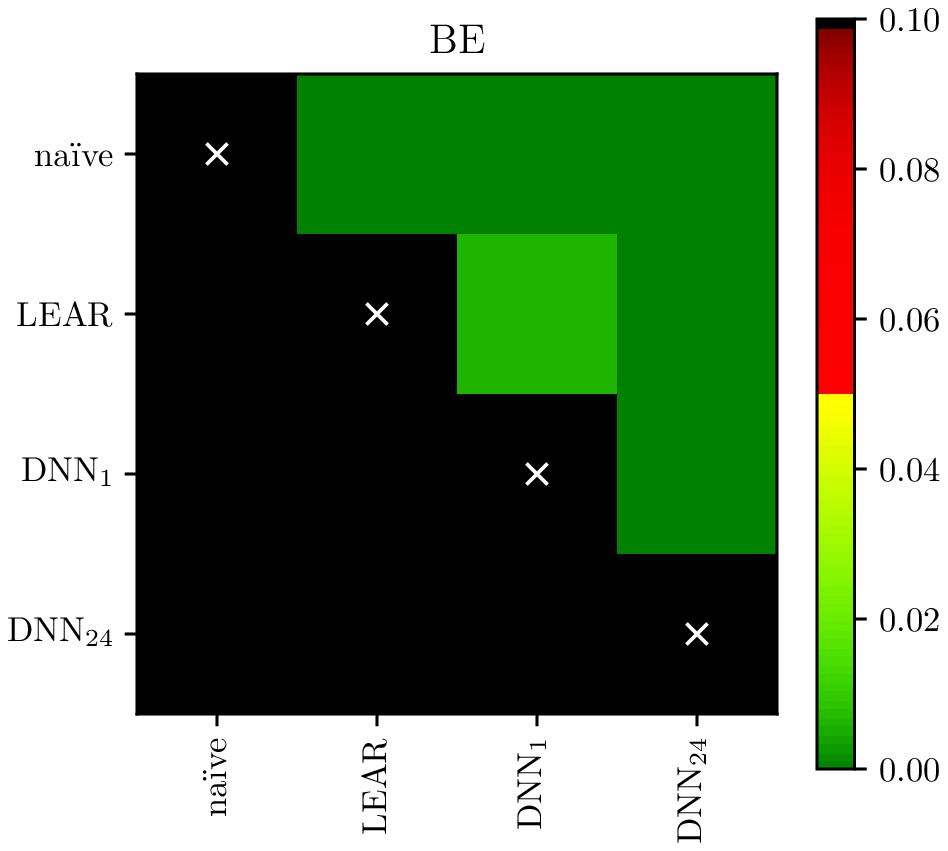}
% 			\caption{EPEX-BE}
		\end{center}
	\end{subfigure}
	\begin{subfigure}[t]{0.16\textwidth}~~~\end{subfigure}
	\begin{subfigure}[t]{0.32\textwidth}
		\begin{center}
			\includegraphics[width=\textwidth]{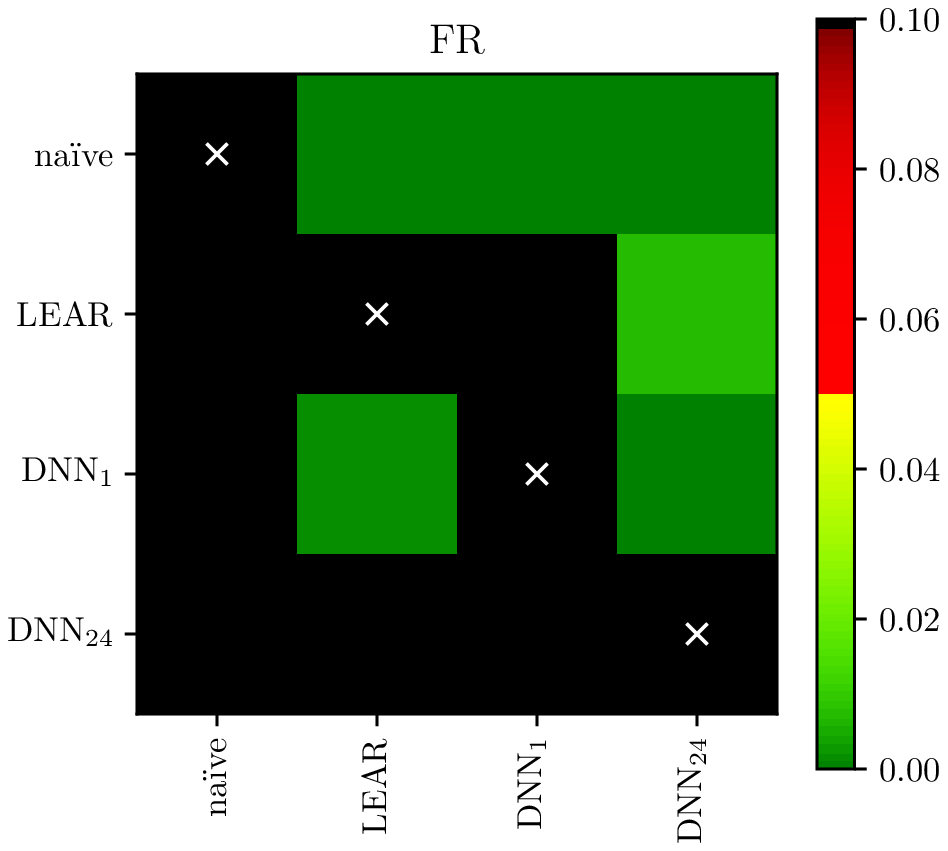}
% 			\caption{EPEX-FR}
		\end{center}
	\end{subfigure}
	\begin{subfigure}[t]{0.32\textwidth}
		\begin{center}
			\includegraphics[width=\textwidth]{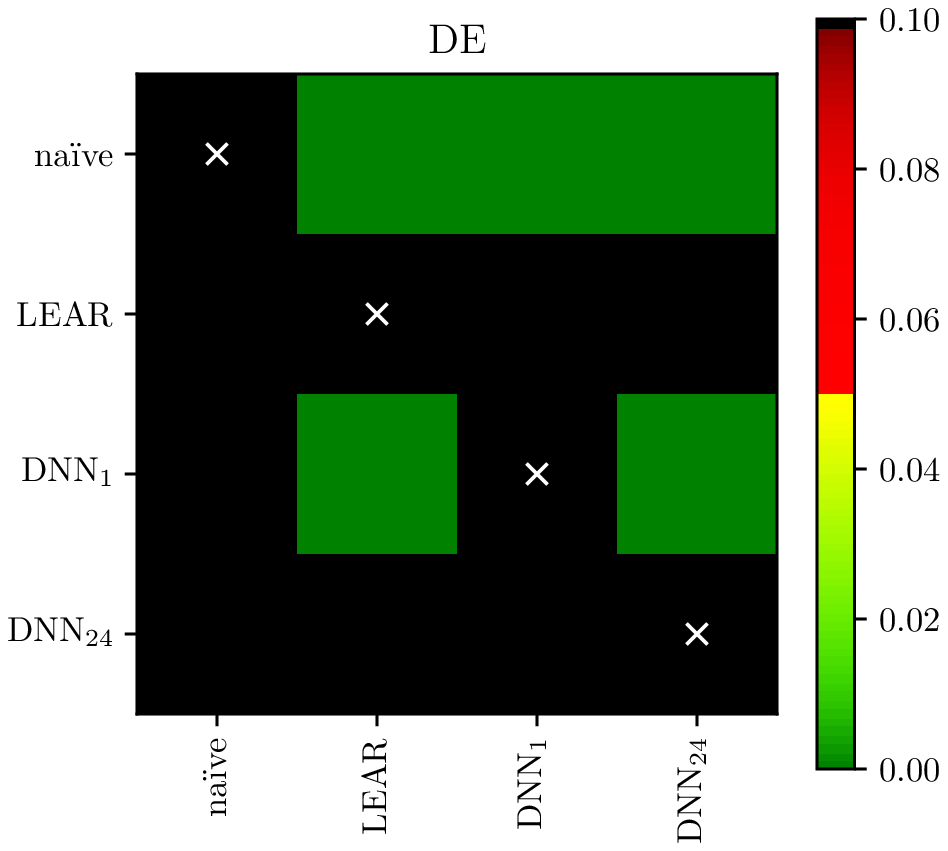}
% 			\caption{EPEX-DE}
		\end{center}
	\end{subfigure}
\caption{Results of the multivariate GW test for each of the five markets. The closer the p-values are to zero (dark green), the more significant the difference is between the forecasts of a model on the X-axis (better) and the forecasts of a model on the Y-axis (worse). Black color indicates p-values above the color map limit, i.e.\ p-values larger or equal than 0.10.}
\label{fig:CPA}
\end{figure*}

Lastly, the outcomes of the models were compared using the Giacomini-White CPA test (see Section \ref{sssec:CPA}). The test results are presented in the form of the chessboards, on which a green square reflects the statistically significant better predictive performance of the model on the x-axis over the one on the y-axis. On the other hand, the black square corresponds to the lack of such an outperformance. Fig. \ref{fig:CPA} consists of 5 panels, each corresponding to one of the datasets.

The immediate conclusion is that the DNN$_{24}$ model is not outperformed by any other model for any of the test datasets. The LEAR and DNN$_1$ models show a comparable performance across, and all of the approaches significantly outperform the na\"ive forecast for all 5 datasets.

\section{Conclusions}
\label{sec:Conclusions}
Using a deep neural network (DNN) model to forecast the electricity prices is not straightforward. An important design aspect involves the structure of the DNN itself: while there are two common structures -- one that focuses on each hour of the day separately, and one that reflects the daily auction structure and models vectors of the prices -- it is not clear the advantages of each architecture. 

This study showed that a DNN methodology that mimicks the LEAR model, i.e., forecasting 24 independent models using single-output networks, might lead to accuracy and computation issues. Although such a setting allows for a fully automatic model specifically designed for each hour of the day (including 24 separate sets of features selected and optimal hyperparameters), for most of the scenarios such an approach is less accurate than the vectorized approach, which due to the multi-output structure is less prone to overfitting errors and can generalize better.
% For some test cases (Nord Pool dataset except hour 4), such an approach exhibits a very solid performance. It would be, therefore, a viable option to consider if the over-fitting issues were reliably avoided.
In particular, the DNN with vectorized outputs shows a solid, robust performance for different  datasets and it is the recommended approach for similar forecasting problems. Additionally, as an extra advantage, it is much simpler (uses only one set of hyperparameters) and computationally less complex -- a single, bigger model is much faster to train than 24 smaller ones.

\bibliography{bibtex/epf}

\begin{thebibliography}{10}
\expandafter\ifx\csname url\endcsname\relax
  \def\url#1{\texttt{#1}}\fi
\expandafter\ifx\csname urlprefix\endcsname\relax\def\urlprefix{URL }\fi
\expandafter\ifx\csname href\endcsname\relax
  \def\href#1#2{#2} \def\path#1{#1}\fi

\bibitem{gre:vas:10}
R.~Green, N.~Vasilakos, Market behaviour with large amounts of intermittent
  generation, Energy Policy 38~(7) (2010) 3211 -- 3220, large-scale wind power
  in electricity markets with Regular Papers.
\newblock \href {http://dx.doi.org/https://doi.org/10.1016/j.enpol.2009.07.038}
  {\path{doi:https://doi.org/10.1016/j.enpol.2009.07.038}}.

\bibitem{bra:bri:hod:16}
C.~{Brancucci Martinez-Anido}, G.~Brinkman, B.-M. Hodge, The impact of wind
  power on electricity prices, Renewable Energy 94 (2016) 474 -- 487.
\newblock \href
  {http://dx.doi.org/https://doi.org/10.1016/j.renene.2016.03.053}
  {\path{doi:https://doi.org/10.1016/j.renene.2016.03.053}}.

\bibitem{nar:zie:19}
M.~Narajewski, F.~Ziel, Econometric modelling and forecasting of intraday
  electricity prices, Journal of Commodity Markets~(DOI:
  10.1016/j.jcomm.2019.100107).

\bibitem{mar:uni:wer:20}
G.~Marcjasz, B.~Uniejewski, R.~Weron, Beating the na{\"\i}ve—combining lasso
  with na{\"\i}ve intraday electricity price forecasts, Energies 13~(7) (2020)
  1667.

\bibitem{epex:18}
EPEX, Annual Report 2017, 2018, http://www.epexspot.com.

\bibitem{wer:14}
R.~Weron, Electricity price forecasting: {A} review of the state-of-the-art
  with a look into the future, International Journal of Forecasting 30~(4)
  (2014) 1030--1081.

\bibitem{mar:uni:wer:19}
G.~Marcjasz, B.~Uniejewski, R.~Weron, On the importance of the long-term
  seasonal component in day-ahead electricity price forecasting with {NARX}
  neural networks, International Journal of Forecasting 35~(4) (2019)
  1520--1532.
\newblock \href {http://dx.doi.org/10.1016/j.ijforecast.2017.11.009}
  {\path{doi:10.1016/j.ijforecast.2017.11.009}}.

\bibitem{lag:rid:sch:18}
J.~Lago, F.~D. Ridder, B.~D. Schutter, Forecasting spot electricity prices:
  Deep learning approaches and empirical comparison of traditional algorithms,
  Applied Energy 221 (2018) 386 -- 405.

\bibitem{wan:zha:che:16}
L.~{Wang}, Z.~{Zhang}, J.~{Chen}, Short-term electricity price forecasting with
  stacked denoising autoencoders, IEEE Transactions on Power Systems 32~(4)
  (2017) 2673--2681.

\bibitem{uni:now:wer:16}
B.~Uniejewski, J.~Nowotarski, R.~Weron, Automated variable selection and
  shrinkage for day-ahead electricity price forecasting, Energies 9 (2016) 621.

\bibitem{zie:wer:18}
F.~Ziel, R.~Weron, Day-ahead electricity price forecasting with
  high-dimensional structures: {U}nivariate vs. multivariate modeling
  frameworks, Energy Economics 70 (2018) 396--420.

\bibitem{Lago2020b}
J.~Lago, G.~Marcjasz, B.~{De Schutter}, R.~Weron, Forecasting day-ahead
  electricity prices: {A} review of state-of-the-art algorithms, best practices
  and an open-access benchmark, Renewable and Sustainable Energy Reviews.

\bibitem{dia:pla:16}
G.~Diaz, E.~Planas, A note on the normalization of {S}panish electricity spot
  prices, IEEE Transactions on Power Systems 31~(3) (2016) 2499--2500.

\bibitem{uni:wer:zie:18}
B.~{Uniejewski}, R.~{Weron}, F.~{Ziel}, Variance stabilizing transformations
  for electricity spot price forecasting, IEEE Transactions on Power Systems
  33~(2) (2018) 2219--2229.
\newblock \href {http://dx.doi.org/10.1109/TPWRS.2017.2734563}
  {\path{doi:10.1109/TPWRS.2017.2734563}}.

\bibitem{now:wer:16}
J.~Nowotarski, R.~Weron, On the importance of the long-term seasonal component
  in day-ahead electricity price forecasting, Energy Economics 57 (2016)
  228--235.

\bibitem{lis:pel:18}
F.~Lisi, M.~Pelagatti, Component estimation for electricity market data:
  Deterministic or stochastic?, Energy Economics 74 (2018) 13--37.
\newblock \href {http://dx.doi.org/10.1016/j.eneco.2018.05.027}
  {\path{doi:10.1016/j.eneco.2018.05.027}}.

\bibitem{hub:mar:wer:19}
K.~Hubicka, G.~Marcjasz, R.~Weron, A note on averaging day-ahead electricity
  price forecasts across calibration windows, IEEE Transactions on Sustainable
  Energy 10~(1) (2019) 321--323.

\bibitem{mar:ser:wer:18}
G.~Marcjasz, T.~Serafin, R.~Weron, Selection of calibration windows for
  day-ahead electricity price forecasting, Energies 11~(9).

\bibitem{naz:far:hei:sha:cat:18}
M.~S. Nazar, A.~E. Fard, A.~Heidari, M.~Shafie-khah, J.~P. Catal{\~{a}}o,
  Hybrid model using three-stage algorithm for simultaneous load and price
  forecasting, Electric Power Systems Research 165 (2018) 214--228.
\newblock \href {http://dx.doi.org/10.1016/j.epsr.2018.09.004}
  {\path{doi:10.1016/j.epsr.2018.09.004}}.

\bibitem{zha:zha:li:tan:ji:19}
J.-L. Zhang, Y.-J. Zhang, D.-Z. Li, Z.-F. Tan, J.-F. Ji, Forecasting day-ahead
  electricity prices using a new integrated model, International Journal of
  Electrical Power {\&} Energy Systems 105 (2019) 541--548.
\newblock \href {http://dx.doi.org/10.1016/j.ijepes.2018.08.025}
  {\path{doi:10.1016/j.ijepes.2018.08.025}}.

\bibitem{benchmarkwebsite}
\href{https://github.com/jeslago/epftoolbox}{Epftoolbox library}.
\newline\urlprefix\url{https://github.com/jeslago/epftoolbox}

\bibitem{epftoolboxdoc}
\href{https://epftoolbox.readthedocs.io}{Epftoolbox documentation}.
\newline\urlprefix\url{https://epftoolbox.readthedocs.io}

\bibitem{Lago2018}
J.~Lago, F.~{De Ridder}, P.~Vrancx, B.~{De Schutter}, Forecasting day-ahead
  electricity prices in {Europe}: The importance of considering market
  integration, Applied Energy 211 (2018) 890--903.
\newblock \href {http://dx.doi.org/10.1016/j.apenergy.2017.11.098}
  {\path{doi:10.1016/j.apenergy.2017.11.098}}.

\bibitem{gia:whi:06}
R.~Giacomini, H.~White, Tests of conditional predictive ability, Econometrica
  74~(6) (2006) 1545--1578.

\bibitem{die:mar:95}
F.~X. Diebold, R.~S. Mariano, Comparing predictive accuracy, Journal of
  Business and Economic Statistics 13 (1995) 253--263.

\bibitem{Bergstra2011}
J.~Bergstra, R.~Bardenet, Y.~Bengio, B.~K{\'{e}}gl, Algorithms for
  hyper-parameter optimization, in: Advances in Neural Information Processing
  Systems, 2011, pp. 2546--2554.

\bibitem{keras}
F.~Chollet, et~al., \href{https://keras.io}{Keras} (2015).
\newline\urlprefix\url{https://keras.io}

\bibitem{kin:ba:14}
D.~P. {Kingma}, J.~{Ba}, {Adam: A Method for Stochastic Optimization}, arXiv
  e-prints (2014) arXiv:1412.6980\href {http://arxiv.org/abs/1412.6980}
  {\path{arXiv:1412.6980}}.

\bibitem{tib:96}
R.~Tibshirani, Regression shrinkage and selection via the lasso, Journal of the
  Royal Statistical Society B 58 (1996) 267--288.

\end{thebibliography}

\end{document}